# INTEGRATION OF AN RFID READER TO A WIRELESS SENSOR NETWORK AND ITS USE TO IDENTIFY AN INDIVIDUAL CARRYING RFID TAGS

Bolivar Torres[1], Qing Pang[2], Gordon W. Skelton[2], Scott Bridges[2], Natarajan Meghanathan[2]

[1]Polytechnic University of Puerto Rico, [2]Jackson State University
[2]natarajan.meghanathan@jsums.edu

***ABSTRACT***

*The objective of this research is to integrate an RFID (Radio Frequency Identification) reader into a Wireless Sensor Network (WSN) to authorize or keep track of people carrying RFID tags. The objective was accomplished by integrating hardware and software. The hardware consisted of two WSN nodes – the RFID node connected to one of the WSN nodes, and a computer connected to the other WSN node. For the RFID equipment, we used the SM130-EK kit, which included the RFID reader and the RFID tags; and for the WSN, we used the Synapse Network Evaluation kit, which included the two sensor nodes. The software consisted of a program module developed in Python to control the microprocessors of the nodes; and a database controlled by a simple program to manage the tag IDs of people wearing them. The WSN and RFID nodes were connected through $I^2C$ interfacing. Also, the work of sending commands to the RFID node, to make it read a tag and send it back to the computer, was accomplished by the Python code developed which also controls the data signals. At the computer, the received tag ID is evaluated with other existing tag IDs on the database, to check if that tag has authorization or not to be in the covered area. Our research has the potential of being adapted for use with secure real-time access control applications involving WSN and RFID technologies.*

## 1. INTRODUCTION

Today's communication technologies promise wide possibilities including Wireless Sensor Networks (WSN), which is a collection of nodes with sensors integrated to collect physical information, such as temperature, pressure, or motion. The sensor nodes consist of processing capabilities (one or more microcontrollers, CPU or DSP chips), may contain multiple types of memory, (program, data and flash memory), and have a power source [1]. These nodes act independently, transmitting the sensed data to a base station for further examination. Nodes can also transmit data to other close nodes, if the base station is far away from that node. This is called multi-hop communication and could improve the energy-cost [4]. Each node contains one or more sensors attached. The type of sensors attached depends on the required application that this node is assigned to do. The WSN technology has the potential of turning the Internet into a physical network [3]. Therefore, WSNs provide a variety of uses for real-time potential applications. Some types of these applications include military, medical, environmental, entertainment and security.

Some of the most important and promising applications on WSNs are in the security field. For example, if one goes to a building or place that requires constant security (bank, corporation, college campus, etc.) it is likely that one will find an access control device, to monitor who enters or goes out from that place, and/or to give them access, or deny it. Radio Frequency Identification (RFID) is the technology of effective automatic identification of different types of objects. The most important functionality of RFID is the ability to track the location of the tagged item [2]. These objects contain attached an RFID tag, which contains the required electronics to be detected by an RFID reader board. This kind of technology is normally seen on places where constant access control and identification is required to get access to the place. RFID is generally composed of three basic components: tags, readers, and PC readers [5]. The tags are attached to any object, with the purpose of being detected by the readers, which will then send the recollected data of the tag to a PC reader that evaluates the information of the tag. The tags can be either active or passive. Active tags require power source to send and receive data, while passive tags do not require power, because they are read-only. For this research, we are using passive tags.

The research described in this paper involves integrating the RFID and WSN technologies, to design a real-time access control application. The basic purpose of this application is to validate the entry of a person entering a protected area to be monitored. A person will be carrying an RFID tag, and when passing through the RFID field he/she will be identified, and will be granted or denied access to the area after validating the read tag ID with the entries in a database. The rest of the paper is organized as follows: Section 2 describes the hardware components used in this research. Section 3 describes the software methods, components and programs used to develop our application. Section 4 illustrates the results obtained from this application software. Section 5 discusses the lessons learnt; Section 6 discusses related work on integration of RFID and wireless sensor networks. Section 7 concludes the paper and presents suggestions for future improvement to the application developed. The terms ‘RFID reader’ and ‘RFID node’ are used interchangeably throughout the paper. They mean the same.

## 2. HARDWARE DEVELOPMENT

The different types of hardware components that were used included the following:

(i) SNAP SN171 Proto Board – This board consists of an RF Engine, which is the core for the radio technology that gives sense to a wireless communication. These boards have input/output capabilities, and are also $I^2C$ (Inter-integrated Circuit) interfacing compatible. Figure 1 shows the SNAP Proto Board. For this research, we are using two of these boards. One will be used as the Polling node, and the other as the RFID node. The Polling node works to make our system able to detect a tag automatically over a certain time period; the RFID node sends and receives signals from the RFID node, each time it is told so by the application software.

(ii) SNAP SN163 Bridge Board – This board also has the RF radio, to communicate with other nodes. The special thing about this node is that on the WSN it acts as the sink, or as an intermediary node communicating with the PC and the other sensor nodes. Thus, this board is essential to transfer any data that reaches it from other nodes, to send them to the PC. We will identify this node as the Bridge node. Figure 2 shows the SNAP Bridge Board. To make things clear, the SNAP nodes, independent of the type, are already programmed from factory, so they can work together. In other words, the connection between these devices is already established and is a ready-to-use system.

(iii) SonMicro SM130 Mifare Read/Write Module – This is the RFID node that reads data from tags close to the magnetic field created by the antenna of this device. It provides types of interfacing such as UART (Universal Asynchronous Receiver/ Transmitter) and $I^2C$ (Inter-integrated Circuit). Figure 3 shows the SonMicro SM130 Mifare Read/Write Module.

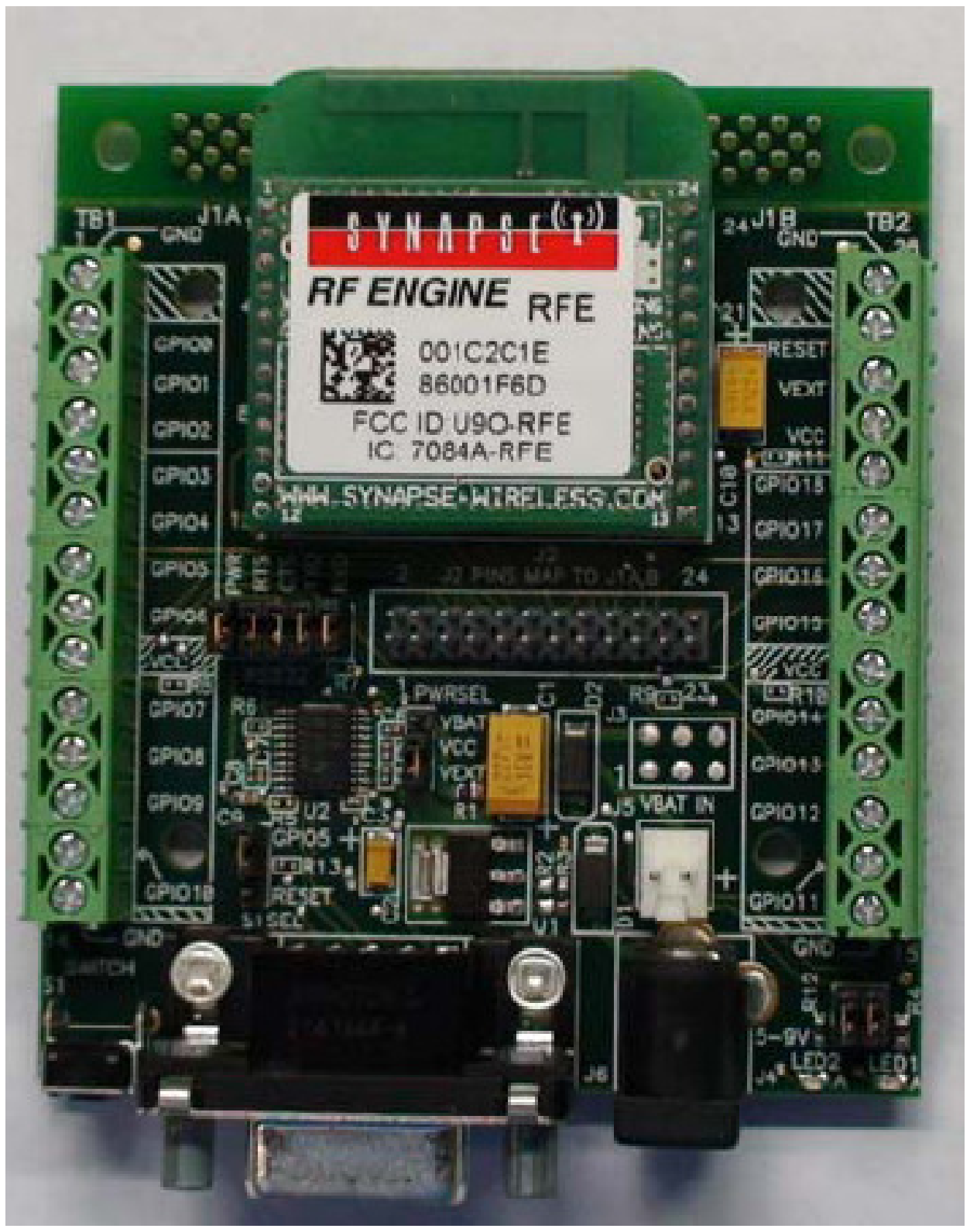

**Figure 1:** SN171 Proto Board

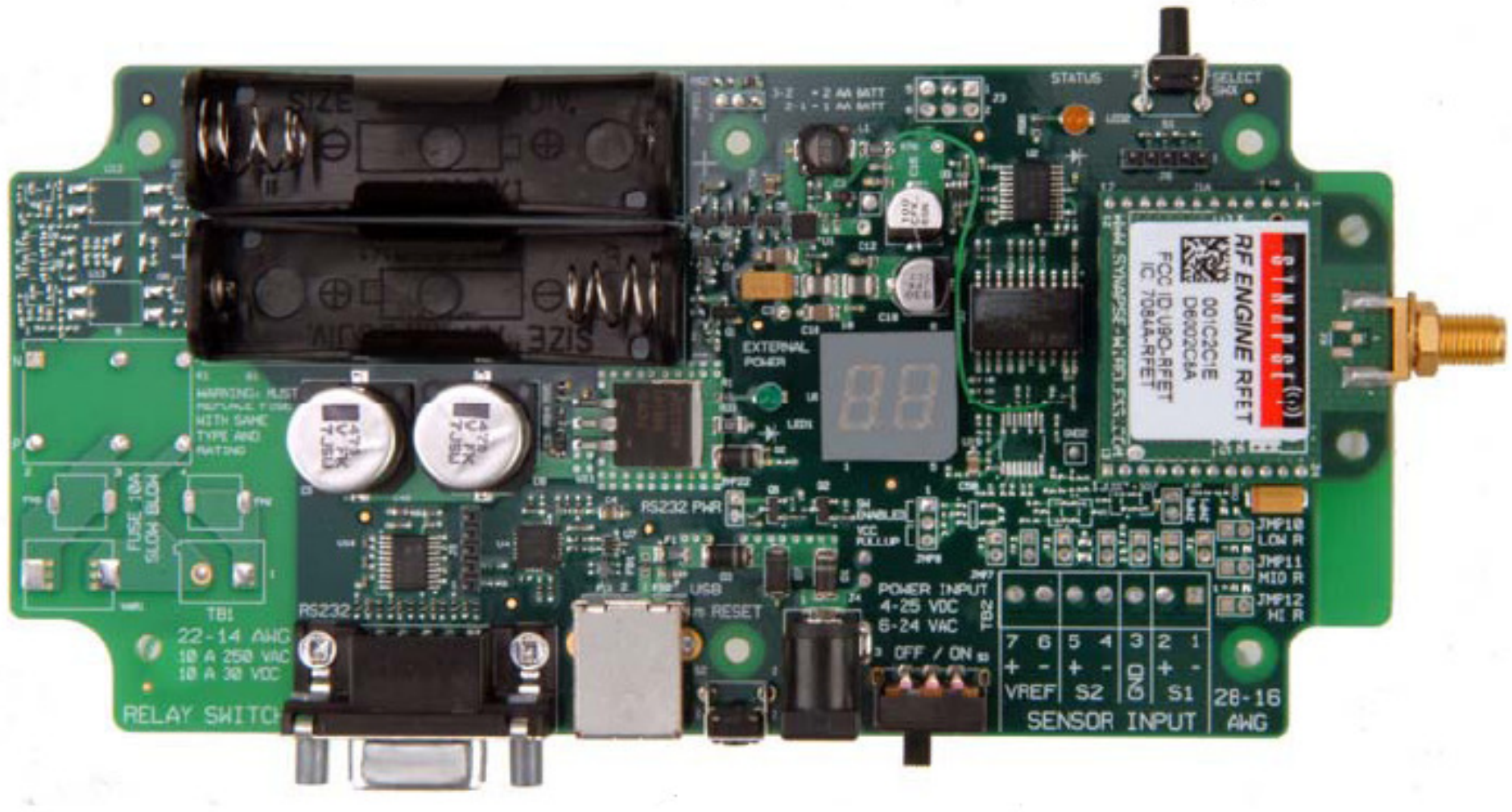

**Figure 2:** SN163 Bridge Board

The interior, almost-flat, part is the antenna of the device. This antenna is capable of detecting tags at least 3 inches away. This device is connected to the RFID reader of the WSN created by the SNAP nodes. For the connection between the WSN node and the RFID reader, we decided to use $I^2C$ interfacing. It was done this way, because both of these devices were $I^2C$ compatible, and the SNAP nodes provide a very useful set of built-in functions to manipulate this kind of interfacing efficiently, and in a simple way.

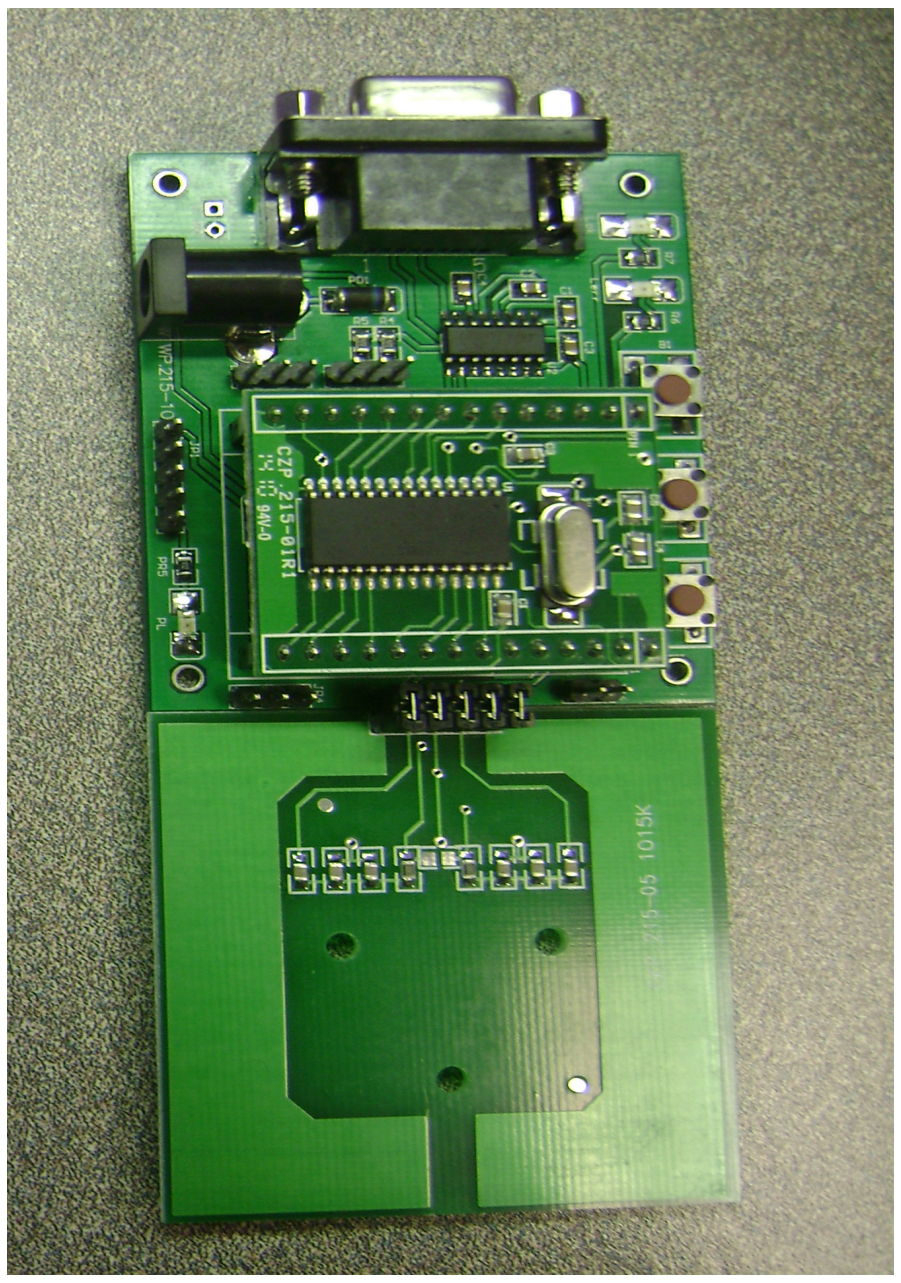

**Figure 3:** SM130 Mifare Read/ Write Module

$I^2C$ interfacing can send/receive parallel data through a serial data line (SDA) and another bus line for the serial clock line (SCL) [6]. The two devices (SNAP node and RFID reader) provide the necessary components and functions to support these two kinds of interfacing. The SNAP node must sequentially send some commands to the RFID reader to control it. The sequence of the data frame for $I^2C$ interfacing for the SM130 Mifare Read/Write Module is specific and has been pre-programmed on the device by factory [7]. Table 1 shows the format of the command frame to be sent from the node through $I^2C$ interfacing.

**Table 1:** Command Frame for $I^2C$ Interfacing

| **Length** | **Command** | **Data** | **CSUM** |
|---|---|---|---|
| 1 Byte | 1 Byte | N Byte | 1 Byte |

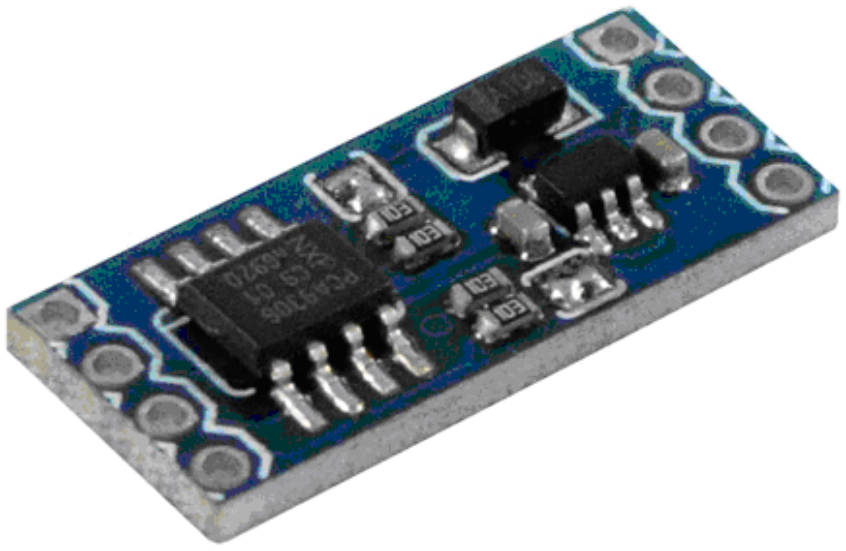

**Figure 4:** $I^2C$/SMBus Voltage Translator (Front)

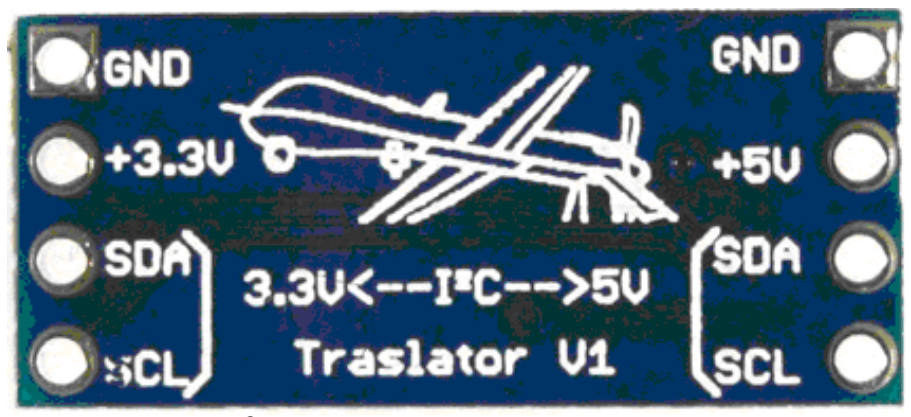

**Figure 5:** $I^2C$/SMBus Voltage Translator (Back)

(iv) $I^2C$/SMBus Voltage Translator ($I^2C$ Level Shifter) – This is a small integrated circuit that allows to interface 3.3V $I^2C$ devices to 5V $I^2C$ devices. We realized that we required this, when we measured the voltages levels at each two of the lines on each board (SCL and SDA). This integrated circuit came very handy because it allowed us to shift the voltage levels on each of the transmitting devices so that they could understand each other. Thus, after installing this tiny circuit, we solved the problem of the connection between the WSN node and the RFID node. Figures 4 and 5 show the front and back view of the voltage translator. The back view helps to understand the logic of the connection of the boards. In this case, the WSN node (SNAP Proto Board) is the one that has a voltage level of 3.3V; while the RFID node (SonMicro SM130 Mifare Read/Write Module) is the one with 5V level operation.

(v) Personal Computer (PC) – This is used to manage the data received and to update the scripts that are running on each node's microcontroller. Also, it serves as the connection to the database part, where the tags are validated. The PC also acts as a node on the WSN and will have a script running on it to manage the data and send it to the database.

## 3. SOFTWARE DEVELOPMENT

The program modules were developed using the Python language. Python is a general-purpose high-level programming language whose design emphasizes code readability. Python is often used as a scripting language for web applications. In our research, we use Python for two potential reasons: to create script modules that will run on the SNAP nodes and to create a program that supports database management [8].

The SNAP nodes can support only a limited number of functions and features from the programming language, because of its limited memory capacity. Therefore, the designers of the SNAP nodes created their own derivation of the Python language called Snappy. This programming language provides the user with the basics of the Python language, such as the usual conditional statements, functions, and logical operators. It also supports three types of variables: Boolean, integers and strings. Also, Snappy provides some built-in functions that are already integrated on the node microcontrollers [9]. These functions provide different possibilities for use with the SNAP nodes, such as doing Remote Procedure Calls (RPC) from one node to another or many nodes. The RPC function is the most relevant, because it is the one that allowed us to communicate the nodes and transmit data, from one node to another. The prototype of the RPC function is like: RPC (address, function, args…). The address field is a unique hex value, for each SNAP node. The function argument is the function to be called on the node of the specified address. The args argument corresponds to the arguments of the function to be called, if there are any required. Also, some of the other important built-in functions are the $I^2C$ interfacing functions. These functions are the following:

- i2cInit(*enablePullups*) – Initializes specific ports on the SNAP node, so that they can be used for $I^2C$ interfacing. If the argument value for *enablePullups* is true, this will enable the pull-up resistors required for the clock and data lines. Nevertheless, the value for this argument is false, because the pullup resistors are already included on the RFID reader.

- i2cWrite(*byteStr, retries, ignoreFirstAck*) – Sends the necessary addressing for the device to respond; and the data frame necessary to execute the desired command (*byteStr*). Argument *retries* makes it possible to retry the number of times specified to make the device to respond if not successful for the first time. Argument *ignoreFirstAck* is true for devices that do not send an initial acknowledgement response. The function returns how much data was sent to the device.

- i2cRead(*byteStr, numToRead, retries, ignoreFirstAck*) – Sends the necessary addressing necessary for the device to respond (*byteStr*). Argument *numToRead* specifies how much data the WSN node is supposed to read from the RFID reader. Argument *retries* makes it possible to retry the number of times specified to make the device to respond if not successful for the first time. Argument *ignoreFirstAck* is true for devices that do not send initial acknowledgement response. The function returns a string with the resulting data, depending on which the command was executed.

Although Snappy is enough to control the working of the SNAP nodes, a graphical interface on the PC is required to write and update the scripts. Synapse Portal (Figure 6) is a standalone program that can run on any PC with Windows 2000 or higher. It uses a USB or RS-232 interface to connect to any of the RF Engines that each SNAP node contains. Thus, this program is basically a user interface to control the network and the software on each of its node. Also, Synapse Portal is the one that makes the PC to act as a node, and implicitly includes it into the network. The SNAP nodes have been integrated with microcontrollers and flash memory so that the scripts written using Snappy can be loaded on them through the Portal and the nodes can have the desired behaviour.

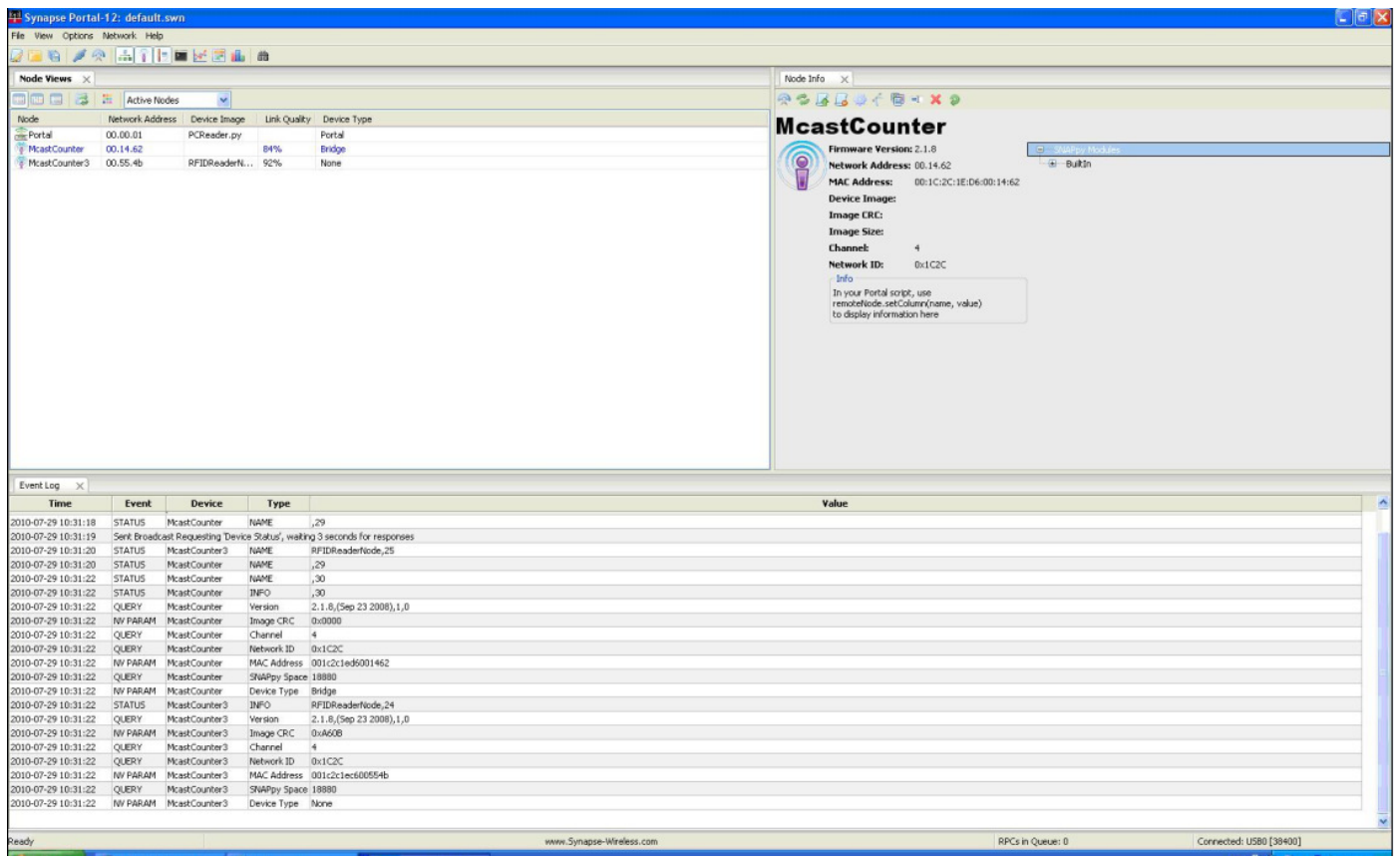

**Figure 6:** Screenshot of Synapse Portal

With all of the above hardware and software components, we could connect and enable communication between these equipment and software such that the tags detected on the RFID node can reach the computer as data for further evaluation and validation with the database tag values. Figure 7 demonstrates the schematic data flow (sequence of steps) in our application.

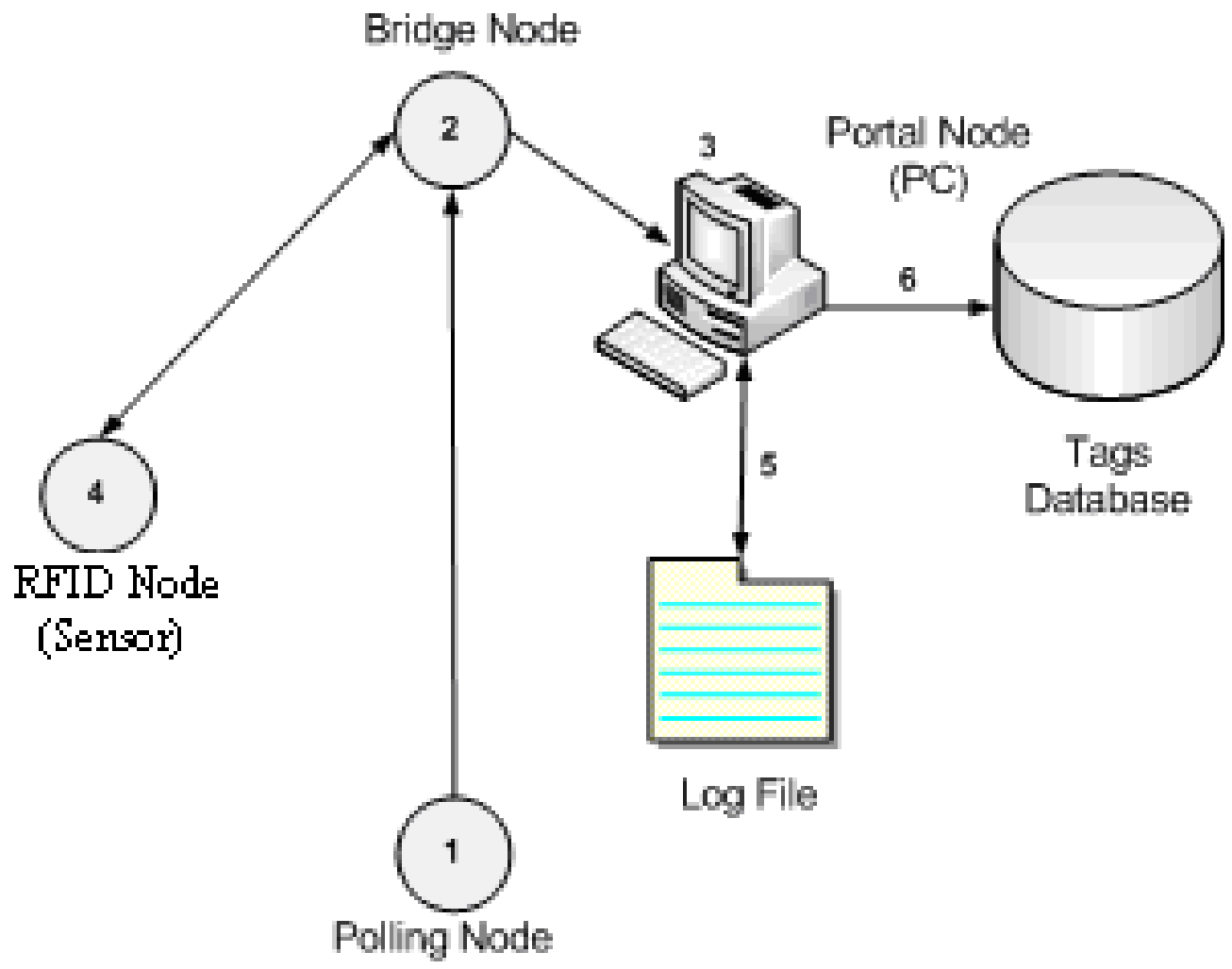

**Figure 7:** Schematic of Data Management

The following are the sequence of steps that the software application takes to continuously detect a tag ID and verify it with the database. Notice that each number in Figure 7 corresponds to each one of these steps, and also indicating where they take place. Steps 1 to 5 can be regarded for 'Node Control' and Step 6 can be regarded for 'Database Control'.

1. The node that starts the procedure of continuously detecting a tag near the RFID node is the Polling node. This node sends an RPC (Remote Procedure Call) for every 2 seconds to the Bridge node and calls the *ping* function, located on the Bridge node. We were first using 5 seconds of poll delay, but after doing further testing, we found that it could be reduced to 2 seconds, for faster detection. Also, there are some additional functions on the script that provide power efficiency and a non-interrupted operation [5]. The following code snippet (Figure 8) is the function inside the Polling node that accomplishes the RPC:

```
def poll100ms():
    #print "Sending Ping"
    rpc('\x00\x14\x62', 'ping')
    #Go back to sleep
    sleep(1, 2)
```

**Figure 8:** Code Snippet for the RPC at the Polling Node

2. Once the *ping* function is called on the Bridge node, it consequently sends an RPC to the PC node to trigger the function that prepares the command to be sent to the RFID node. The ping function on the Bridge board is simple and consists of the code snippet shown in Figure 9.

```
def ping():
    rpc('\x00\x00\x01', 'selectTag')
```

**Figure 9:** Code Snippet for the *ping* function at the Bridge Node

3. The PC node receives the signal from the RPC issued on the Bridge node and executes a function that builds the command frame necessary for the RFID node to detect a tag. The command frame is then sent to the RFID node through a RPC. The code snippet shown in Figure 10 demonstrates how this is accomplished. Notice that one of the arguments for the RPC issued from the PC node to the RFID node has a *callback* function argument (a built-in function of Snappy) and it specifies that function to invoke with the return value of the remote function. In other words, we use the *callback* to get back the result from the RFID node into a function defined on the PC node.

```
def selectTag( ):
  global currentCmd
  i2cFrame = ""
  i2cFrame += chr(slaveAddr)
  i2cFrame += chr(0x01)#Length
  i2cFrame += chr(0x83)#Command
  i2cFrame += chr(0x84)#CSUM
  cmdSent = "selectTag"
  currentCmd = cmdSent
  sendCommand(i2cFrame, cmdSent)
```

```
def sendCommand(i2cFrame, cmdSent):
  rpc('\x00\x55\x4B', 'callback',
      'receiveResult',
      'sendReceiveCommand', slaveAddr,
      i2cFrame, cmdSent)
```

**Figure 10:** Code Snippet for Building and Sending the Command Frame from the PC to the RFID Node

4. After the RFID node receives the ping from the PC node, it executes a function that writes to the RFID node the command, using the $I^2C$ functions. The result is then read from the RFID node using the same function and returned to the PC node. This is accomplished through the code snippet shown in Figure 11 function is the following code snippet (Comments begin with '#' and are shown here for readability):

```
def sendReceiveCommand(slaveAddr, i2cFrame, cmdSent):
  cmd = i2cFrame
  #Send command frame to RFID reader.
  bytesReturned = i2cWrite(cmd, 1, True)
  #Do an OR bitwise operation of the slave address value with 1. Required to read frame from RFID node
  returnAddr = chr(slaveAddr | 1)
  #Read response from RFID reader.
  dataReturned = i2cRead(returnAddr, frameSize, 4, True)
  return bin2hex(dataReturned)
```

**Figure 11:** Code Snippet for Using the $I^2C$ functions at the RFID Node

```
def receiveResult(data):
  print data
  if currentCmd == "selectTag":
    i = 0
    j = 0
    found = False
    cleanData = ""
    cleanData += "0"
    while data[i] == '0':#Find the start of the received frame
      i += 1
    packageSize = int(data[i])

#Sum 2 to include the length and the CSUM, and multiply by 2 to be sure we take the entire byte
    while j < ((packageSize + 1) * 2) - 1:
      cleanData += data[i]
      i += 1
      j += 1
    if packageSize != 2: #Tag present on field
      found = True
      now = datetime.now( )
      dateTimeNoMs = str(now.day) + '-'
      dateTimeNoMs += str(now.month) + '-'
      dateTimeNoMs += str(now.year) + ' '
      dateTimeNoMs += str(now.hour) + ':'
      dateTimeNoMs += str(now.minute) + ':'
      dateTimeNoMs += str(now.second)
      dateTimeNoMs += '\t'
      message = dateTimeNoMs
      message += cleanData[6:]
              #Trim the length, command, and tagType bytes
      message += '\n'
      writeToFile(message)#Call function to acccess file and write result
      appendToOut(dateTimeNoMs + data)
      return found
  else:
    writeToFile(data)
```

**Figure 12:** Synapse Portal Code Snippet Running at the PC Node to Process the Data Frame returned from the RFID Node

5. After the data frame containing the tag has been retrieved from the RFID node, the data is returned to the Synapse portal function (code snippet, with comments marked using #, is shown in Figure 12) running on the PC that takes the data containing the tag and trims it of the other unnecessary data. The 'select tag' command from the RFID node does not only return the tag ID itself, but also data such as the length of the command and data, the command itself and the checksum. After trimming the unnecessary data, we save the tag on a log file; along with the date and time it was detected, to make sure that if the same tag is constantly detected, we know how to differentiate them.

6. While the Polling node is constantly sending RPCs to continuously detect the tag IDs, our application has another independent program (developed in Python) that runs and detects new tag IDs written on the log file, with a delay of 2 seconds, to process and visualize the results. This program reads the log file, and if there is an unchecked tag ID in the log, it will be retrieved and validated with the entries in a database of tag IDs, created with Microsoft

Office Access (MS Access). We used the Pyocdb libraries for connecting the database to the program. The various fields in the MS Access database are shown in Table 2.

**Table 2:** Fields in the MS Access Database of Tags

| Tag Fields |
|---|
| Tag ID (Primary Key) |
| Is Authorized |
| First Name |

```
while i < logSize:
    #Compare list sizes, to check if each tag detected was verified
    curFrame = curFList[i].split('\t')
    if len(curFrame) == 2:
      found = True
      break
    i += 1
if found == True:
  #Tag found on RF field
  cnxn = pyodbc.connect('DRIVER={Microsoft Access Driver (*.mdb,
  *.accdb)};DBQ=C:\\Documents and Settings\\btorres\\My
  Documents\\DBSupport\\RFID_System_I2C.accdb;')
  cursor = cnxn.cursor( )
  cursor.execute("SELECT Tags.[Tag ID], Tags.[First Name], Tags.[Is Authorized] FROM Tags
  WHERE Tags.[Tag ID] = '" + newTag + "';")
  if len(retrievedRow) != 0:
    #Tag found on database.
    ...
    if retrievedRow[0][2] == 'Yes':
        #Access granted.
        ...
    else:
        #Access denied
        ...
```

**Figure 13:** Code Snippet to Validate the Tag IDs with the Entries in the MS Access Database

After verifying the unchecked tag ID value with the database, the program will write on the log either of these three values along the tag ID and the date: ‘Y’ (the person was found, and is authorized), ‘N’ (the person was found, but is not authorized) or “NF” (tag ID not found on database). Figure 13 is a code snippet that shows the relevant conditions of the program and the query executed to retrieve data from the database.

The program to access the database was developed with a Graphical User Interface (GUI) to enable the user to input the running time of the program, which should ideally run on an infinite loop. For testing and research purposes, we adopted the approach of running the validation program for specific time period (in seconds) that would be input through the GUI, illustrated in Figure 14.

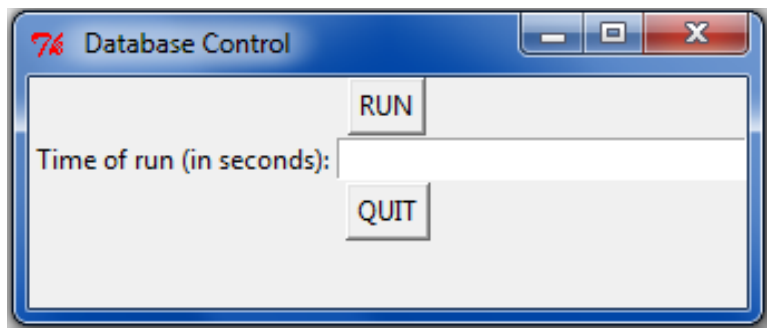

**Figure 14:** Database Control Graphical User Interface

# 4. EXPERIMENT RESULTS

The data provided in this section was collected from a series of tests, conducted by running the two parts of the software (Nodes Control & Database Control) simultaneously to prove that they can run and write the log file at the same time. The Nodes Control software will always be running during the tests, while the Database Control software will run for the specified quantity of seconds. For these tests, we assume that a tag is always present in the RFID field. Figures 15 through 18 illustrate the observations from the two tests with a polling frequency of 5 seconds in each case. While Figures 15 and 16 respectively show the output generated by the Synapse Portal script for the data frame retrieved from the RFID device and Database Control for a time period of 60 seconds; Figures 17 and 18 depict the results generated when the tests were run for 30 seconds.

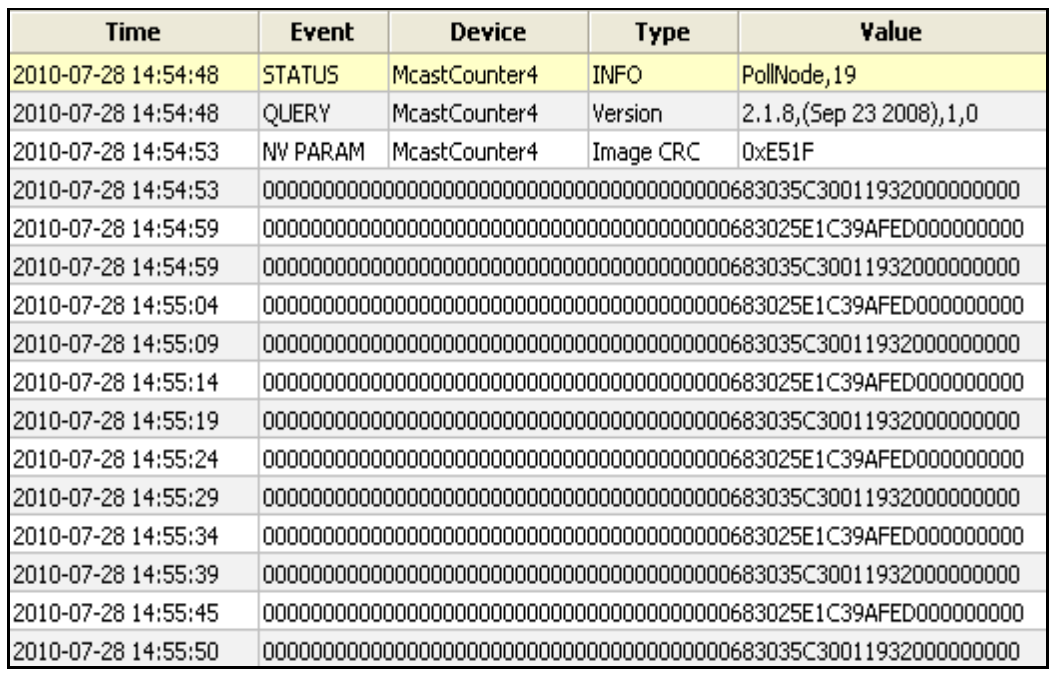

| Time | Event | Device | Type | Value |
|---|---|---|---|---|
| 2010-07-28 14:54:48 | STATUS | McastCounter4 | INFO | PollNode,19 |
| 2010-07-28 14:54:48 | QUERY | McastCounter4 | Version | 2.1.8,(Sep 23 2008),1,0 |
| 2010-07-28 14:54:53 | NV PARAM | McastCounter4 | Image CRC | 0xE51F |
| 2010-07-28 14:54:53 | 00000000000000000000000000000000000000683035C3001193200000000 | | | |
| 2010-07-28 14:54:59 | 00000000000000000000000000000000000000683025E1C39AFED00000000 | | | |
| 2010-07-28 14:54:59 | 00000000000000000000000000000000000000683035C3001193200000000 | | | |
| 2010-07-28 14:55:04 | 00000000000000000000000000000000000000683025E1C39AFED00000000 | | | |
| 2010-07-28 14:55:09 | 00000000000000000000000000000000000000683035C3001193200000000 | | | |
| 2010-07-28 14:55:14 | 00000000000000000000000000000000000000683025E1C39AFED00000000 | | | |
| 2010-07-28 14:55:19 | 00000000000000000000000000000000000000683035C3001193200000000 | | | |
| 2010-07-28 14:55:24 | 00000000000000000000000000000000000000683025E1C39AFED00000000 | | | |
| 2010-07-28 14:55:29 | 00000000000000000000000000000000000000683035C3001193200000000 | | | |
| 2010-07-28 14:55:34 | 00000000000000000000000000000000000000683025E1C39AFED00000000 | | | |
| 2010-07-28 14:55:39 | 00000000000000000000000000000000000000683035C3001193200000000 | | | |
| 2010-07-28 14:55:45 | 00000000000000000000000000000000000000683025E1C39AFED00000000 | | | |
| 2010-07-28 14:55:50 | 00000000000000000000000000000000000000683035C3001193200000000 | | | |

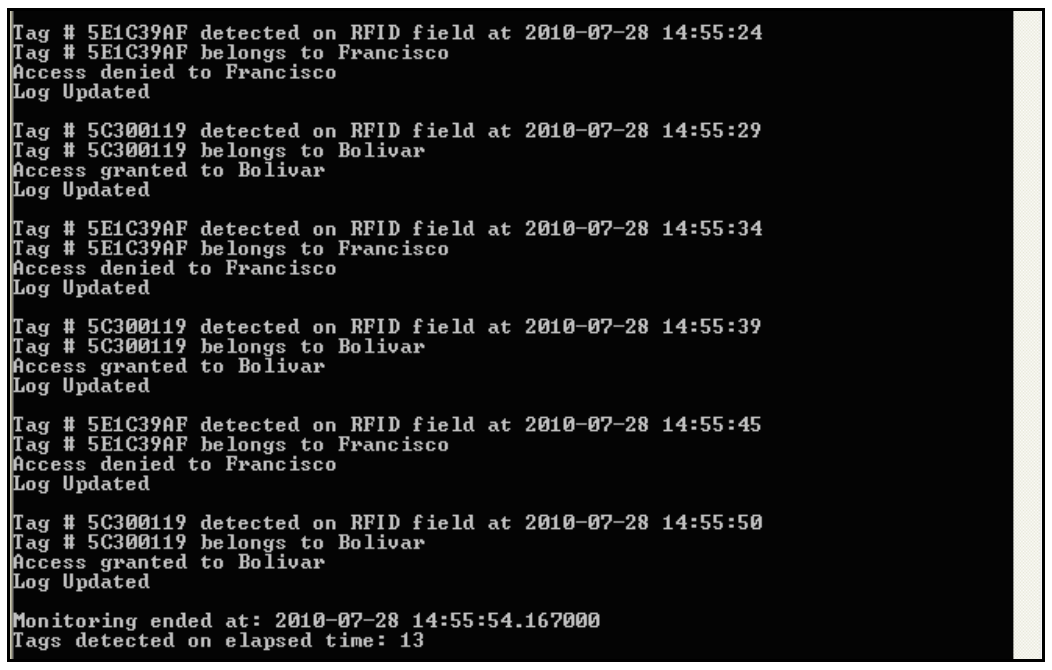

**Figure 15:** Portal Log Output for 60 Seconds **Figure 16:** Database Control for 60 Seconds

Besides the tests shown in Figures 15 through 18, we also performed two other tests, with a polling delay of 2 seconds (as shown in Table 3), to check whether our application provides a reliable running time without stopping. Because of space constraints, the screenshots of the running of these two 2-second polling delay tests are not shown. However, we summarize the results – running time, number of tags detected, starting time and ending time – of all these four tests in Table 3. Table 4 shows the percentage of error that the results in Table 3 incur, compared to the theoretical values of the number of tags that were supposed to be detected with the given time conditions (polling delay and run time). The theoretical value for the number of tags that could be detected is calculated by dividing the run time by the polling delay.

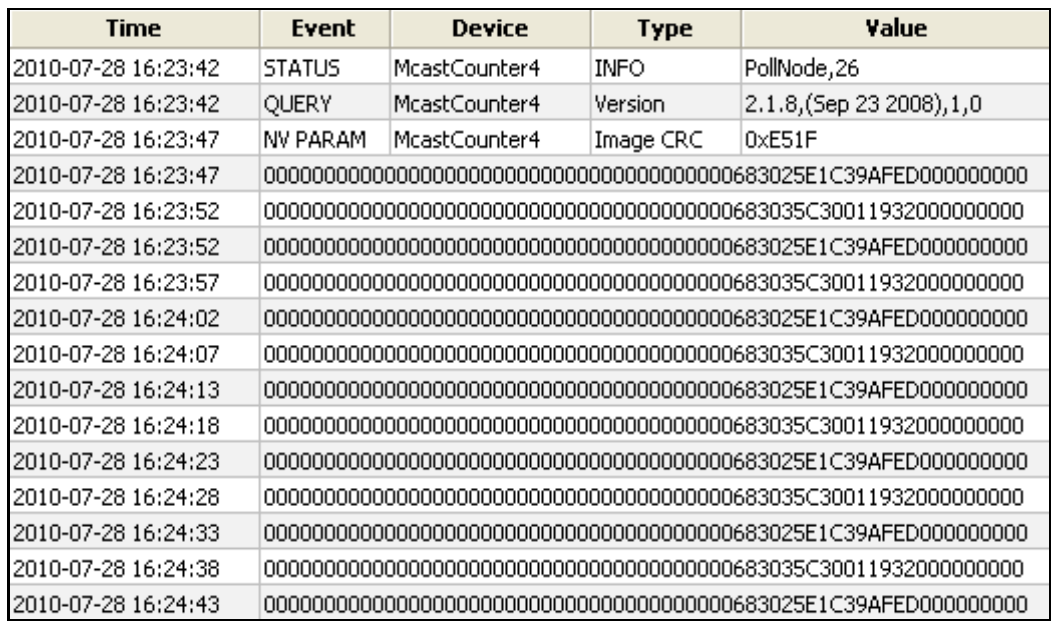

**Figure 17:** Portal Log Output for 30 Seconds

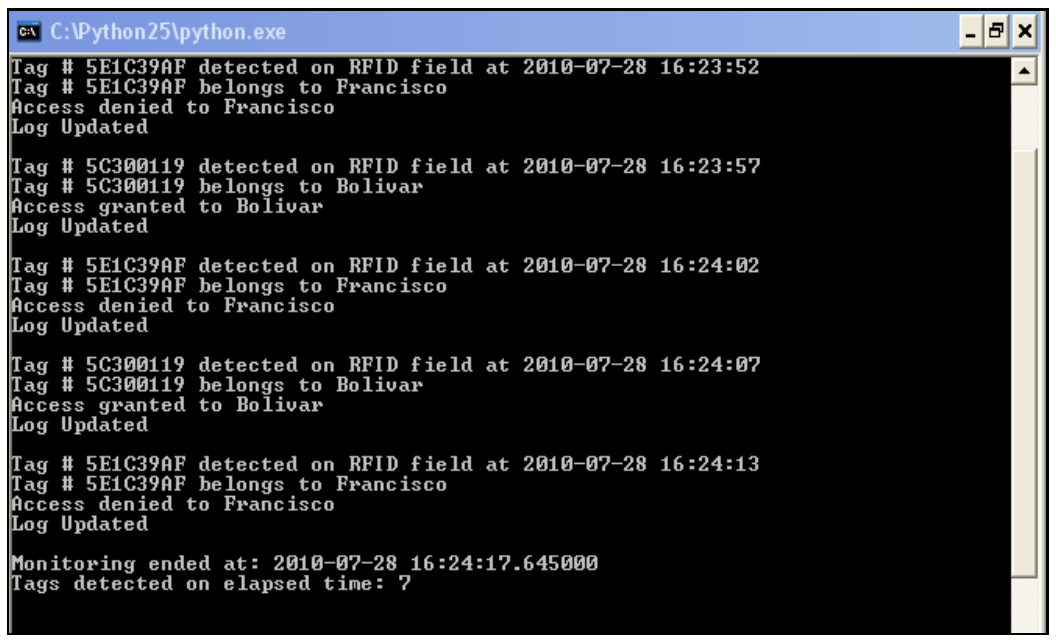

**Figure 18:** Database Control for 30 Seconds

**Table 3:** Tests Performed on different conditions

| Poll Delay (s) | Run Time (s) | Tags Detected | Starting Time | Ending Time |
|---|---|---|---|---|
| 5 | 60 | 13 | 2010-07-28 14:54:54 | 2010-07-28 14:55:54 |
| 5 | 30 | 7 | 2010-07-28 16:23:47 | 2010-07-28 16:24:17 |
| 2 | 120 | 59 | 2010-08-02 14:20:21 | 2010-08-02 14:22:21 |
| 2 | 3600 | 1712 | 2010-08-02 12:17:34 | 2010-08-02 13:17:34 |

**Table 4:** Percentage of Error Generated: Theoretical vs. Experimental Results

| Theoretical Tags Detected | Experimental Tags Detected | Percentage of Error (%) |
|---|---|---|
| 12 | 13 | 8.33 |
| 6 | 7 | 16.17 |
| 60 | 59 | 1.67 |
| 1800 | 1712 | 4.89 |
| **Average Percentage of Error** | 7.77% | |

## 5. LESSONS LEARNT

The objectives established for this research were accomplished. Through the methods used on the development of the access control application, we understood that integrating WSN and RFID technologies can be more complicated than the expected, and especially when we are trying to design a real-time application, which involves both hardware and software. Normally a WSN integrates resistive-type sensors (accelerometers, thermistors, photocells, etc), but in this case we are using an actual RFID device, which requires low-level programming knowledge to adapt it to the WSN successfully. Through the process, we noticed that the hardware at first did not present the requirements for the application to be done. Thus, we researched looking for devices such as the RFID reader, and the $I^2C$ voltage translator, so that we could set up all the hardware, before trying any fully testing with the software. In other words, we realized that when dealing with hardware/software, the hardware part must be the first to be fully completed before starting on writing any code or script modules for the SNAP nodes, or the PC. The software part was very distributed, in a way that individual scripts were written to be then uploaded to the SNAP nodes, and to also run them on the computer, using the portal, or simply the Python terminal. Thus, integrating all this software to make it work as a single unit was a work that required understanding every detail of the code. For example, the RPC functions were basically the ones that made the SNAP nodes to communicate through each other. These RPC functions required specific arguments, and especially on the address to which node it will be doing the RPC, because not every SNAP node acted the same way, with each one of them performed a different task.

## 6. RELATED WORK

The following works have been reported in the literature regarding integration of RFID to wireless sensor networks:

(i) In [10], the authors propose a prototype for an in-home health care system to monitor the medication in take of patients. The prototype comprised of two Mica2 motes [11] – a RFID reader mote and a base station mote – used for radio communication, a RFID reader simulator communicating via a serial port to the reader mote and a personal computer as the base station, also connected to the base station mote via a serial port.

(ii) Intel, Inc., has developed a Wireless Identification and Sensing Platform (WISP) [12] that comprises of passive RFID tags that gather their operating energy from Ultra High Frequency (UHF) RFID reader transmissions and also include sensors (e.g., accelerometers) that provide a very small-scale computing platform. WISPs are a viable alternative to motes for smart dust applications owing to their small form factor and lack of battery. A pilot study on efficiently using WISPs through a dense sensing approach for monitoring and recognizing human indoor activities has been reported in [13].

(iii) A general-purpose RFID tag (referred to as S-tag) has been proposed in [14] for connecting to a generic sensor and transmitting, when interrogated, the measured physical parameter. S-tags are very cost-effective as they use UHF for communication. S-tags can be implemented using either a "multi-chip" strategy – with RFID tags consisting of only one antenna and many chips, each with its own ID code or a "multi-tag" strategy – with a different tag for transmitting each ID, controlled by appropriate microwave circuits.

(iv) A RFID-based sensor, to alert firefighters within minutes of a fire ignition [15], has been designed by Telepathx, a wireless and communications company based in Melbourne, Australia. The proposed Variable Radio Frequency (VRF) sensor comprises of an active RFID tag and wireless thermal sensors that activate the tag when the temperature of the sensed region exceeds 2 degrees of a pre-determined setting. The RFID tag, when activated, communicates its identity to a reader that in turn sends notification to the fire fighters to respond quickly and efficiently.

(v) A prototype for a child localization system using RFID and wireless sensor network technologies has been designed and implemented in [16]. The proposed system uses passive RFID tags coupled with motion detection sensors to locate children in the park and convey the information to a remote personal computer (base station) through UHF RFID readers. The system can even identify multiple children visiting a site together.

(vi) In [17], the authors present a Non-deterministic Pushdown Automata (NPA) model to RFID transceivers with wireless sensor networks for tracking and monitoring animals. Here, the sensor nodes integrated with the RFID readers operate with two read distances: one read distance for the RFID reader to gather data from the tags and another read distance for the sensor node to communicate with the peer sensor nodes and form an ad hoc network that can be used for aggregating the gathered data to a remote sink node.

(vii) In [18], the authors present taxonomy for the integration of RFID and wireless sensor networks. Four classes of integration have been envisioned: integrating RFID tags with sensors, integrating RFID tags with wireless sensor network nodes and other wireless devices, integrating RFID readers with sensor nodes and other wireless devices, and a mix of RFID tags, readers and wireless sensor nodes. The authors emphasize the potential benefits in integrating the RFID technology that easily facilitates detection and identification of objects with the wireless sensor network technology that provides information about the condition of the objects, the environment as well as enables multi-hop wireless communications.

## 7. CONCLUSIONS AND FUTURE WORK

Based on the results provided in this paper, we can make some relevant analysis and draw conclusions. The runtime values for the application ranged from 30 seconds to 3600 seconds (1 hour). Each test was conducted to examine whether our application runs for different ranges of time, and explore the constant detection of tags by the RFID reader. The results were fairly accurate based on the average percentage of error analysis. We observe that the largest error percentage generated was 16.17%, which corresponded to the test with polling delay of 5 seconds, and running time of 30 seconds. This value is high for a percentage error, and the conclusion we can take from here is that not only it was the smaller value for the running time that makes the reading inaccurate, but also the accuracy of the procedure to start the test, because each program must be started independently by the user. The other percentages of error were lower and less than 10%. We would also like to comment on the test with a polling delay of 2 seconds and a running time of 3600 seconds (1 hour). Although, this result shows a lower percentage of error, we can see a big difference on the number of experimental tags detected and the theoretical tags detected. This difference may be happening because of the time error. We must remember that although the Node Control and Database Control parts of the application have time delays, they may not be necessarily accurate. Thus, we must also take in consideration the time that each program takes on executing each its own instructions.

Some of the suggestions for future improvements for this application include the following: (i) Make the two parts of the program to run as one, completely independent of the user; (ii) Develop a multiprocessing interface, to make the Database Control program run, and have the option to stop it, change the rate of detection, etc; (iii) Make the SNAP functions and scripts more compatible with Python so that the nodes can work independent of the Synapse Portal; (iv) Add more RFID nodes with its corresponding RFID devices integrated, so the Polling node can do a multicast RPC to all nodes to detect a tag ID; (v) Add accelerometers to the nodes of the WSN to detect movement, if there is no tag present.

## ACKNOWLEDGMENTS

This research and publication are funded by the U.S. National Science Foundation (NSF) through grants CNS-0851646 and DUE-0941959. The student author Bolivar Torres was one of the participants of the Summer 2010 NSF-sponsored Research Experiences for Undergraduates (REU) program on Wireless Ad hoc Networks and Sensor Networks, hosted by the Department of Computer Science at Jackson State University (JSU), MS, USA. The authors also acknowledge Dr. Loretta Moore, Mrs. Brenda Johnson and Ms. Ilin Dasari (all at JSU) for their services to the REU 2010 program.